\documentstyle[twoside]{article}

\catcode`\@=11
\long\def\@makefntext#1{
\protect\noindent \hbox to 3.2pt {\hskip-.9pt  
$^{{\eightrm\@thefnmark}}$\hfil}#1\hfill}		

\def\@makefnmark{\hbox to 0pt{$^{\@thefnmark}$\hss}}	
	
\def\ps@myheadings{\let\@mkboth\@gobbletwo
\def\@oddhead{\hbox{}
\rightmark\hfil\eightrm\thepage}   
\def\@oddfoot{}\def\@evenhead{\eightrm\thepage\hfil
\leftmark\hbox{}}\def\@evenfoot{}
\def\sectionmark##1{}\def\subsectionmark##1{}}

\oddsidemargin=\evensidemargin
\newcounter{sectionc}\newcounter{subsectionc}\newcounter{subsubsectionc}
\renewcommand{\section}[1] {\vspace{12pt}\addtocounter{sectionc}{1} 
\setcounter{subsectionc}{0}\setcounter{subsubsectionc}{0}\noindent 
	{\tenbf\thesectionc. #1}\par\vspace{5pt}}
\renewcommand{\subsection}[1] {\vspace{12pt}\addtocounter{subsectionc}{1} 
	\setcounter{subsubsectionc}{0}\noindent 
	{\bf\thesectionc.\thesubsectionc. {\kern1pt \bfit #1}}\par\vspace{5pt}}
\renewcommand{\subsubsection}[1] {\vspace{12pt}\addtocounter{subsubsectionc}{1}
	\noindent{\tenrm\thesectionc.\thesubsectionc.\thesubsubsectionc.
	{\kern1pt \tenit #1}}\par\vspace{5pt}}
\newcommand{\nonumsection}[1] {\vspace{12pt}\noindent{\tenbf #1}
	\par\vspace{5pt}}

\newcounter{appendixc}
\newcounter{subappendixc}[appendixc]
\newcounter{subsubappendixc}[subappendixc]
\renewcommand{\thesubappendixc}{\Alph{appendixc}.\arabic{subappendixc}}
\renewcommand{\thesubsubappendixc}
	{\Alph{appendixc}.\arabic{subappendixc}.\arabic{subsubappendixc}}

\renewcommand{\appendix}[1] {\vspace{12pt}
        \refstepcounter{appendixc}
        \setcounter{figure}{0}
        \setcounter{table}{0}
        \setcounter{lemma}{0}
        \setcounter{theorem}{0}
        \setcounter{corollary}{0}
        \setcounter{definition}{0}
        \setcounter{equation}{0}
        \renewcommand{\thefigure}{\Alph{appendixc}.\arabic{figure}}
        \renewcommand{\thetable}{\Alph{appendixc}.\arabic{table}}
        \renewcommand{\theappendixc}{\Alph{appendixc}}
        \renewcommand{\thelemma}{\Alph{appendixc}.\arabic{lemma}}
        \renewcommand{\thetheorem}{\Alph{appendixc}.\arabic{theorem}}
        \renewcommand{\thedefinition}{\Alph{appendixc}.\arabic{definition}}
        \renewcommand{\thecorollary}{\Alph{appendixc}.\arabic{corollary}}
        \renewcommand{\theequation}{\Alph{appendixc}.\arabic{equation}}
        \noindent{\tenbf Appendix \theappendixc #1}\par\vspace{5pt}}
\newcommand{\subappendix}[1] {\vspace{12pt}
        \refstepcounter{subappendixc}
        \noindent{\bf Appendix \thesubappendixc. {\kern1pt \bfit #1}}
	\par\vspace{5pt}}
\newcommand{\subsubappendix}[1] {\vspace{12pt}
        \refstepcounter{subsubappendixc}
        \noindent{\rm Appendix \thesubsubappendixc. {\kern1pt \tenit #1}}
	\par\vspace{5pt}}

\newcommand{\textlineskip}{\baselineskip=13pt}
\newcommand{\smalllineskip}{\baselineskip=10pt}

\def\eightcirc{
\begin{picture}(0,0)
\put(4.4,1.8){\circle{6.5}}
\end{picture}}
\def\eightcopyright{\eightcirc\kern2.7pt\hbox{\eightrm c}} 

\newcommand{\copyrightheading}[1]
	{\vspace*{-2.5cm}\smalllineskip{\flushleft
	{\footnotesize Modern Physics Letters A, #1}\\
	{\footnotesize $\eightcopyright$\, World Scientific Publishing
	 Company}\\
	 }}

\def\abstracts#1#2#3{{
	\centering{\begin{minipage}{4.5in}\baselineskip=10pt\footnotesize
	\parindent=0pt #1\par 
	\parindent=15pt #2\par
	\parindent=15pt #3
	\end{minipage}}\par}} 

\newcommand{\bibit}{\nineit}
\newcommand{\bibbf}{\ninebf}
\renewenvironment{thebibliography}[1]
	{\frenchspacing
	 \ninerm\baselineskip=11pt
	 \begin{list}{\arabic{enumi}.}
        {\usecounter{enumi}\setlength{\parsep}{0pt}     
	 \setlength{\leftmargin 12.7pt}{\rightmargin 0pt} 
         \setlength{\itemsep}{0pt} \settowidth
	{\labelwidth}{#1.}\sloppy}}{\end{list}}

\newcounter{itemlistc}
\newcounter{romanlistc}
\newcounter{alphlistc}
\newcounter{arabiclistc}

\newcommand{\fcaption}[1]{
        \refstepcounter{figure}
        \setbox\@tempboxa = \hbox{\footnotesize Fig.~\thefigure. #1}
        \ifdim \wd\@tempboxa > 5in
           {\begin{center}
        \parbox{5in}{\footnotesize\smalllineskip Fig.~\thefigure. #1}
            \end{center}}
        \else
             {\begin{center}
             {\footnotesize Fig.~\thefigure. #1}
              \end{center}}
        \fi}

\newcommand{\tcaption}[1]{
        \refstepcounter{table}
        \setbox\@tempboxa = \hbox{\footnotesize Table~\thetable. #1}
        \ifdim \wd\@tempboxa > 5in
           {\begin{center}
        \parbox{5in}{\footnotesize\smalllineskip Table~\thetable. #1}
            \end{center}}
        \else
             {\begin{center}
             {\footnotesize Table~\thetable. #1}
              \end{center}}
        \fi}

\def\@citex[#1]#2{\if@filesw\immediate\write\@auxout
	{\string\citation{#2}}\fi
\def\@citea{}\@cite{\@for\@citeb:=#2\do
	{\@citea\def\@citea{,}\@ifundefined
	{b@\@citeb}{{\bf ?}\@warning
	{Citation `\@citeb' on page \thepage \space undefined}}
	{\csname b@\@citeb\endcsname}}}{#1}}

\newif\if@cghi
\def\cite{\@cghitrue\@ifnextchar [{\@tempswatrue
	\@citex}{\@tempswafalse\@citex[]}}
\def\citelow{\@cghifalse\@ifnextchar [{\@tempswatrue
	\@citex}{\@tempswafalse\@citex[]}}
\def\@cite#1#2{{$\null^{#1}$\if@tempswa\typeout
	{IJCGA warning: optional citation argument 
	ignored: `#2'} \fi}}

\def\@refcitex[#1]#2{\if@filesw\immediate\write\@auxout
	{\string\citation{#2}}\fi
\def\@citea{}\@refcite{\@for\@citeb:=#2\do
	{\@citea\def\@citea{, }\@ifundefined
	{b@\@citeb}{{\bf ?}\@warning
	{Citation `\@citeb' on page \thepage \space undefined}}
	\hbox{\csname b@\@citeb\endcsname}}}{#1}}

\def\@refcite#1#2{{#1\if@tempswa\typeout
        {IJCGA warning: optional citation argument
	ignored: `#2'} \fi}}

\def\refcite{\@ifnextchar[{\@tempswatrue
	\@refcitex}{\@tempswafalse\@refcitex[]}}

\def\pmb#1{\setbox0=\hbox{#1}
	\kern-.025em\copy0\kern-\wd0
	\kern.05em\copy0\kern-\wd0
	\kern-.025em\raise.0433em\box0}

\def\fnt#1#2{\footnotetext{\kern-.3em
	{$^{\mbox{\scriptsize #1}}$}{#2}}}

\def\fpage#1{\begingroup
\voffset=.3in
\thispagestyle{empty}\begin{table}[b]\centerline{\footnotesize #1}
	\end{table}\endgroup}

\def\runninghead#1#2{\pagestyle{myheadings}
\markboth{{\protect\footnotesize\it{\quad #1}}\hfill}
{\hfill{\protect\footnotesize\it{#2\quad}}}}
\font\tenrm=cmr10
\font\tenit=cmti10 
\font\tenbf=cmbx10
\font\bfit=cmbxti10 at 10pt
\font\ninerm=cmr9
\font\nineit=cmti9
\font\ninebf=cmbx9
\font\eightrm=cmr8

\def\qed{\hbox{${\vcenter{\vbox{			
   \hrule height 0.4pt\hbox{\vrule width 0.4pt height 6pt
   \kern5pt\vrule width 0.4pt}\hrule height 0.4pt}}}$}}

\begin{document}
\runninghead{M. Losanow-Kirchbach}{Balmer-like Series for Baryon $\ldots$}
\normalsize\textlineskip
\thispagestyle{empty}
\setcounter{page}{1}
\copyrightheading{}
\vspace*{0.88truein}

\fpage{1}

\centerline{\bf BALMER-LIKE SERIES FOR BARYON RESONANCES}

\vspace*{0.21truein}

\centerline{\footnotesize M. LOSANOW-KIRCHBACH
\footnote{E-mail: mariana@kph.uni-mainz.de}}

\centerline{\footnotesize
Institut f\"ur Kernphysik, J. Gutenberg Universit\"at, D-55099 Mainz, Germany}

\vspace*{0.21truein}

\hspace{4.7cm}In memory of my first teacher on group

\hspace{4.7cm}theory, the Bulgarian physicist

\hspace{4.7cm}Angel Nicolov.

\vspace*{0.21truein}

\abstracts{
The pole positions of various baryon resonances have been found
to reveal a well pronounced clustering, the so-called H\"ohler cluster.
In a previous work, the H\"ohler clusters have been shown to be
identical to Lorentz multiplets of the type
$\lbrace {1\over 2}+l', {1\over 2}+l'\rbrace
\otimes \lbrack\lbrace {1\over 2} ,0\rbrace \oplus
\lbrace 0,{1\over 2}, 0\rbrace \rbrack $
with $l'$ integer. Here we show that the cluster positions
are well described by means of a Balmer-series like recursive
mass formula.}{}{}

\vspace*{0.25truein}
\noindent
One of the oldest unsolved problems of baryon spectroscopy is the relativistic
description of resonances with higher-spin. This problem appears both
in treating the higher-spin states as pointlike objects 
as well as at the composite particle level in the various quark models.
In the first case the problem is caused by the presence of redunant
components, that need be eliminated, in the multi-spinor representations 
of the Lorentz group embedding the higher-spin states considered.
In the second case, the problem is mainly
due to the improper choice for the symmetry group of the constituent
quark model. 

As known, baryons have been considered so far 
to constitute multiplets of the group O(3)$\otimes$SU(6)$_{SF}$.
Through this group the trivial spin-flavor $(SF)$ correlation
between three quarks in the 1s-shell is naively extended to arbitrary 
orbital angular momenta. In this way the relativistic description of the
higher-spin states is completely spoiled. Indeed, the Lorentzian boost
breaks up the spin-flavor correlation. It couples the spin to different
O(3) representations and produces mixing between the 
O(3)$\otimes$SU(6)$_{SF}$ multiplets. 
The problems raised by symmetry groups of strong interaction
based on strong correlations between the spin- and flavor 
degrees of freedom of quarks don't have only 
conceptual disadvantages. Also their predictions are not
satisfactory. For example, the  O(3)$\otimes$SU(6)$_{SF}$
classification scheme predicts numerous still unobserved 
higher-spin states known as `missing resonances'.
Finally, the spacing among the corresponding multiplets is much smaller as 
compared to the mass splitting between the multiplet members.
On the other side, 
a recent analysis of the pole positions of various baryon 
resonances  ($L_{2I, 2J}$) with masses below $\sim $ 2500 MeV performed 
by H\"ohler {\em et al.} [\refcite{Hoehler}] 
reveals a well-pronounced clustering.
This is quite a surprising result in as it was not 
anticipated by any model or theory [\refcite{Nefkens}].
In the following, these baryon clusters will be referred to as
`H\"ohler clusters' according to a suggestion of Nefkens [\refcite{Nefkens}].
In conjunction with this observation, the symmetry of all reported 
$N$, $\Delta $ and $\Lambda $ 
baryon excitations with masses below 2500 MeV was re-analyzed 
[\refcite{Ki1,Ki2}] 
and shown to be governed by SL(2,C) $\otimes $SU(2)$_I$. As long as the 
group SL(2,C) is the universal covering of the Lorentz group, the new 
classification scheme for baryons is  determined by O(1,3)$\otimes $SU(2)$_I$ 
rather than by O(3)$\otimes $ SU(6)$_{SF}$ as always used 
since the invention of the naive three flavor quark model.
The O(1,3)$\otimes $SU(2)$_I$ symmetry indicates that
the spin-orbital correlation between quarks is much stronger than the 
spin-flavor one. Indeed, in Refs. [\refcite{Ki1,Ki2}] it was demonstrated 
that H\"ohler's clusters are identical to Lorentz multiplets of the
type $\lbrace {1\over 2}+l' ,{1\over 2}+l' \rbrace \otimes 
\lbrack \lbrace 1/2,0\rbrace \oplus \lbrace 0, 1/2\rbrace\rbrack $
with $l' $ integer. These representations can be interpreted to
decribe a quark coupled to Di-quarks of (integer) spins 
$l=0,..., 2l'+1 $ as emerging through the coupling of two spin-1/2 
objects occupying an $l'$ shell.
Indeed, the covariant quark-Di-quark model based on solving the
Bethe-Salpeter equation reveals an internal O(4) symmetry
as visible from the rapid convergence of its solutions in the basis
of the Gegenbauer polynomials, the orthogonal polynomials of the
group O(4), an observation reported in Refs. [\refcite{Kus}].

The Lorentz multiplets introduced above are
well known from the Coulomb problem, where they
correspond to even principal quantum numbers $n=  2l'+2$.
In the following, these multiplets will be sometimes termed as
Coulomb-spinors.
All orbital angular momenta contained within a Coulomb multiplet
have either natural or unnatural parity. Coupling a Dirac spinor to $l$ 
is then standard and leads to states with total spins $\vec{J}=\vec{l}
\otimes \vec {1\over 2}$ with $l=0, ..., 2l'+1$. The parity $(-1)^{L+1}$ of 
a single $\pi N$ resonance $L_{2I,2J}$ in standard notation \cite{Part} is 
determined in the present classification scheme by either $(-1)^l$ or 
$(-1)^{l+1}$, depending on whether the parity of the intrinsic orbital angular 
momentum is natural or unnatural. In the present notation, L takes the values of 
either $L=|l-1|, (l+1)$ for natural, or $L=l$ for unnatural parities.

\begin{table}[htbp]
\tcaption{Correspondence between H\"ohler clusters and Lorentz multiplets.
 The five predicted missing resonances here have been labeled by `$ ms $'. 
 }
\vspace*{0.21truein}
\begin{tabular}{lll}
\hline
\\
$\bf\mbox{L} _{2I,2J} $ {\bf States} & {\bf Pole (MeV)} & 
{\bf Lorentz Multiplet} \\
\\ 
\hline
~\\
S$ _{11} $, P$ _{11} $, P$ _{13} $, D$ _{13} $, & (1665 $\pm $25) & 
$ \lbrace\frac{3}{2}, \frac{3}{2}\rbrace \otimes $
$ \lbrack \lbrace \frac{1}{2}, 0 \rbrace \oplus $
$ \lbrace 0,\frac{1}{2}\rbrace \rbrack $ \\
D$ _{15} $, F$ _{15} $, F$ _{17}^{ms} $ 
& -(55 $\pm $ 15)i & $ \otimes \lbrace \frac{1}{2}\rbrace _I $ \\
 & & ( n=4)\\
~\\
\hline
~\\
S$ _{11} $, P$ _{11} $, P$_{13}$, D$ _{13} $,  &
(2110 $\pm $ 50) & $ \lbrace \frac{5}{2},\frac{5}{2}\rbrace \otimes $
$ \lbrack \lbrace \frac{1}{2}, 0 \rbrace \oplus \lbrace 0, \frac{1}{2} $
$ \rbrace \rbrack $\\
D$ _{15} $, F$_{15}$, F$ _{17} $, G$ _{17} $, &
-(180 $\pm $50)i & $\otimes \lbrace \frac{1}{2} \rbrace _I $   \\
 G$ _{19} $, H$ _{19} $, H$ _{1,11}^{ms} $ &  & (n=6) \\
~\\
\hline
~\\
S$ _{31} $, P$ _{31} $, P$ _{33} $, D$ _{33} $, & (1820 $\pm $30)&
$ \lbrace \frac{3}{2}, \frac{3}{2}\rbrace \otimes $ 
$ \lbrack \lbrace \frac{1}{2},0 \rbrace \oplus \lbrace 0,\frac{1}{2} $
$\rbrace\rbrack  $\\
D$ _{35} $, F$ _{35} $, F$ _{37} $ & -(120 $\pm $30)i &
$\otimes \lbrace \frac{3}{2} \rbrace_I $  \\
 & & (n=4)\\
~\\
\hline
~\\
S$ _{31} $, P$ _{31}^{ms} $, P$ _{33}^{ms} $, D$_{33}^{ms} $, &  &
$\lbrace \frac{5}{2}, \frac{5}{2} \rbrace \otimes \lbrack $
$ \lbrace \frac{1}{2},0 \rbrace \oplus \lbrace 0,\frac{1}{2} $
$\rbrace\rbrack  $ \\
D$ _{35} $, F$ _{35}$, F$ _{37}$, G$ _{37} $ &
less established & $\otimes \lbrace \frac{3}{2}\rbrace _I $ \\
G$ _{39} $, H$_{39}$, H$ _{3,11} $ & & (n=6) \\
~\\
\hline
~\\
S$ _{01} $(1800) , P$ _{01} $(1810) &  & 
$ \lbrace 0^\pm\rbrace \otimes \lbrack \lbrace \frac{1}{2},0 \rbrace $ 
$ \oplus \lbrace 0,\frac{1}{2}\rbrace \rbrack $\\
D$ _{05} $(1830), F$ _{05} $(1820) &  &
$\lbrace 0^\pm \rbrace \otimes\lbrack \lbrace \frac{5}{2},0 \rbrace $ 
$\oplus \lbrace 0, \frac{5}{2}\rbrace \rbrack $\\
P$ _{03} $ (1890), D$ _{03}^{ms} $(2000?) &  &
$\lbrace 0^\pm\rbrace \otimes\lbrack \lbrace \frac{3}{2},0 \rbrace $ 
$\oplus \lbrace 0,\frac{3}{2}\rbrace \rbrack $\\
G$ _{07} $(2100), F$ _{07} $(2020) &  &
$\lbrace 0^\pm\rbrace \otimes\lbrack \lbrace \frac{7}{2},0 \rbrace $ 
$\oplus \lbrace 0,\frac{7}{2}\rbrace \rbrack $\\
~\\
\hline
\end{tabular}
\end{table}

The new baryon spectrum generating algebra o(1,3)$\otimes $su(2)$_I$
has numerous advantages over the old o(3)$\otimes$su(6)$_{SF}$ one.
For example, the spacing of about
200 MeV among the relativistic multiplets
with masses below 2000 MeV becomes much larger as compared to 
the mass splitting of 50-70 MeV between the corresponding multiplet 
members. Furthermore, fewer  `missing resonances' are predicted. 
For example, within the O(1,3)$\otimes $SU(2)$_I$ scheme, the $\Delta $ 
spectrum below 2 GeV appears complete. As the F$_{37}$ state has to be 
paralleled in the nucleon sector by a (still unobserved) F$_{17}$ resonance 
with a mass around 1700 MeV, only that latter state has to be viewed as a 
`missing resonance' among the non-strange baryon excitations with masses 
below 2000 MeV. In comparing the states from the third nucleon and 
$\Delta $ clusters, four more missing resonances are predicted. 
These are the H$_{1, 11}$, P$_{31}$, P$_{33}$, and D$_{33}$ states with 
masses between 2200 and 2400 MeV. 

In summary, five new, still unobserved non-strange resonances have been 
predicted in Refs. [\refcite{Ki1,Ki2}] (see Table 1).

The major advantage of the relativistic spectrum generating algebra for 
baryons is, however, that it reconciles such seemingly 
contrary ideas of the baryon structure like the constituent quark
model on the one side, and the structureless (pointlike) view of
hadrons, on the other side. Indeed, the Lorentz representations used for 
describing pointlike higher-spin states emerge in a natural way from an 
underlying quark-Di-quark picture of baryon structure. In that case the 
lower-spin components of the Lorentzian multi-spinors embedding the 
higher-spin states are not any longer redundant but correspond to observed 
resonances. As a consequence, the relativistic description 
of higher-spin states becomes possible because the relativistic propagators 
of the H\"ohler clusters can be directly read off from the representation 
theory of the Lorentz group. For example, the ($S_{2I, 1}, D_{2I, 3}$ ) 
cluster is described in terms of a Lorentz vector with Dirac spinor components 
and its propagator is given by (see Ref. [\refcite{Ki2}] for details )
\begin{equation}
S_{\mu\nu} = {{(p^\lambda\gamma_\lambda +M)(g_{\mu\nu}
-{ {p_\mu p_\nu}\over M^2} )}
\over {2M(p^2-M^2)}}\, ,\qquad \mu =0,1,2,3\, ,
\label{cluster_prop}
\end{equation}
with M standing for the mass of the degenerate resonances under 
considerations. In noting that, say, the first S$_{11}$ and D$_{13}$ 
states appear separated by only 15 MeV, one sees that the relativistic 
contribution of these states to the amplitude of processes like meson 
photoproduction at threshold, can easily be calculated. Along the line 
of the  representation theory of the Lorentz group, the construction of
higher-cluster propagators is straightforward.

\begin{table}[htbp]
\tcaption{Predicted ($M_{jj}^{th}$) and reported 
($M_{jj}^{exp}$) positions (in MeV) of the Balmer-like baryon 
lines together with the maximal deviation ($\delta^{max}$) 
of a resonance mass from the cluster mass-average value. 
To keep notations transparent, the Lorentz multiplets have been
represented by the quantum numbers $\lbrace j,j\rbrace _B$ of their
Coulomb multiplet parts only with $B=N,\Delta$ and $\Lambda $.}

\vspace*{0.21truein}

\begin{tabular}{lccc}
\hline
\\
{\bf  $\bf
 \lbrace j,j\rbrace _B$ }  & {\bf $\bf M_{jj}^{th}$ }  & 
{\bf $\bf M_{jj}^{exp}$ } & $\bf \delta^{max} $\\
\\ 
\hline
~\\
$\lbrace {1\over 2},{1\over 2}\rbrace _N$ & 1467 & 1498 &58 \\ \\
$\lbrace {1\over 2}+1,{1\over 2}+1\rbrace _N$& 1734&1689& 31 \\ \\
$\lbrace {1\over 2}+2,{1\over 2}+2 \rbrace _N$&2068&2102& 148\\
~\\
\hline
~\\
$\lbrace {1\over 2},{1\over 2}\rbrace _\Delta $ &
1696 & 1690&70 \\ \\
$\lbrace {1\over 2}+1,{1\over 2}+1\rbrace _\Delta $ &
2005 & 1922&28   \\ \\
$\lbrace {1\over 2}+2,{1\over 2}+2\rbrace _\Delta $ & 
2391& 2276 &144 \\ 
~\\
\hline
~\\
$\lbrace {1\over 2},{1\over 2}\rbrace _\Lambda$ & 
 1605&1508& 103\\
~\\
\hline
~\\
\end{tabular}
\end{table}

Now the apparent analogy between the spectrum of the hydrogen atom
and the baryon spectra leads to the question whether or not
the positions of the Lorentz clusters follow a sort of Balmer-series 
like pattern\footnote{I like to thank J\"org Friedrich for his remark
during one of my talks that the spacing among the H\"ohler clusters 
is as well pronounced as that among the Balmer lines.}.   
The answer to that question is positive. Below quite a simple  
empirical recursive relation is suggested that describes with quite an 
amazing accuracy the reported mass averages of the resonances
from the Lorentz multiplets  only in terms of the cluster quantum numbers 
and the masses of the ground state baryons:
\begin{eqnarray}
{        {M_{ l'+{1\over 2}, l'+{1\over 2}}      }
\over {M_{l'-{1\over 2},l'-{1\over 2} } }      }
&=&1+
\left( {1\over {(2l'+1)^2}}-{1\over {(2l'+2)^2}}\right)
J^{max}_{l'-{1\over 2},l'-{1\over 2}}
\left(J^{max}_{l'-{1\over 2},l'-{1\over 2}} +1\right) \, ,\nonumber\\
J^{max}_{l'-{1\over 2},l'-{1\over 2}} &=& 2l'-{1\over 2}\, , 
\qquad l'>0\, . \label{Balmer_ser}
\end{eqnarray}
Here, M$_{j,j}$ denotes the mass of the respective Lorentz multiplet,
while J$^{max}_{j,j}$ stands for the maximal (half-integer) spin of the
multiplet $\lbrace j,j\rbrace \otimes \lbrack 
\lbrace {1\over 2},0\rbrace \oplus \lbrace 0,{1\over 2}\rbrace \rbrack$.
The position of the first excited nucleon cluster with $l'=0$ is
related to the nucleon mass $M_N$ via
\begin{eqnarray}
M_{{1\over 2}, {1\over 2}} &=& M_N + (1-{1\over 4})\, {1\over 2}
\, ({1\over 2}+1)M_N\, .
\label{first_line}
\end{eqnarray}
For the $\Delta $ and $\Lambda $ baryons, the mass scale
entering Eq.~(\ref{first_line}), has to be replaced by 
$ (M_N+M_\Delta )/2$, and $ (M_N+M_\Lambda )/2$, respectively.
In Table 2, the comparison between the reported mass averages
of the resonances constituting a Lorentz cluster
and the positions of the predicted Balmer-like baryon lines
following from Eq.~(\ref{Balmer_ser}) is given.
The table leads to the insight that the pattern underlying the baryon 
spectrum is that of the `Balmer-like' resonance series. 

The recursive empirical mass formula in Eq.~(\ref{Balmer_ser}) may be
expected to result from a proper effective quark potential having 
O(4) symmetry in the leading order. This O(4) symmetry has to be slightly 
violated if one wishes to explain the mass splitting between the members 
of the Lorentz clusters under consideration, that especially for the 
$\lbrace {1\over 2}, {1\over 2}\rbrace $ and 
$\lbrace {3\over 2},{3\over 2}\rbrace $ representations
turns out to be surprisingly small.
Unearthing such a potential may lead to a further understanding of
the nature of strong interaction and the structure of hadrons.

To conclude we wish to note that while the SU(6)$_{SF}$ classification 
scheme for baryons may still be useful to unify the ground states of  
the spin-1/2 and 3/2 baryons to the respective octet and decuplet, 
and to establish Gell-Mann's mass formulae,
the relativistic O(1,3)$\otimes $SU(2)$_I$ symmetry
is  much better suited for describing the baryon excitations.

\nonumsection{Acknowledgement}
This work was supported by the Deutsche Forschungsgemeinschaft (SFB 201).

\nonumsection{References}

\end{document}